\documentclass[reprint, aps, floatfix, superscriptaddress, numeric, longbibliography, nofootinbib]{revtex4-2}

\usepackage{import}
\usepackage{amsmath}
\usepackage{amssymb}
\usepackage{graphicx}
\usepackage[normalem]{ulem}
\usepackage{dcolumn}
\usepackage{afterpage}
\usepackage{tabularx}
\usepackage{multirow}
\usepackage{booktabs}
\usepackage{hhline}
\usepackage[table]{xcolor}
\definecolor{tableShade}{gray}{0.9}
\usepackage{makecell}
\usepackage{array}
\usepackage{float}
\usepackage{soul}
\usepackage{comment}
\usepackage{bm}
\usepackage[colorlinks=true, linkcolor=blue, citecolor=blue]{hyperref}
\usepackage{orcidlink}

\def\fr#1#2{{{#1} \over {#2}}}
\def\half{{\textstyle{1\over 2}}}
\def\frac#1#2{{\textstyle{{#1}\over {#2}}}}

\def\sb{\overline{s}{}}
\def\re{{\rm Re}~}
\def\im{{\rm Im}~}

\newcommand{\beq}{\begin{equation}}
\newcommand{\eeq}{\end{equation}}
\newcommand{\bea}{\begin{eqnarray}}
\newcommand{\eea}{\end{eqnarray}}

\newcommand{\rf}[1]{(\ref{#1})}

\newcommand{\UWM}{Department of Physics, University of Wisconsin-Milwaukee, Milwaukee, WI 53201, USA}
\newcommand{\CarlPhys}{Department of Physics and Astronomy, Carleton College, Northfield, MN 55057, USA}

\begin{document}

\title{Measuring gravitational wave speed and Lorentz violation with the first three gravitational-wave catalogs}

\author{Anarya Ray\,\orcidlink{0000-0002-7322-4748}}
\email{anarya@uwm.edu}
\affiliation{\UWM}

\author{Pinchen Fan\,\orcidlink{0000-0003-3988-9022}}
\affiliation{\CarlPhys}

\author{Vincent F. He\,\orcidlink{0009-0001-1045-7536}}
\affiliation{\CarlPhys}

\author{Malachy Bloom\,\orcidlink{0000-0001-5489-4204}}
\affiliation{\CarlPhys}

\author{Suyu Michael Yang}
\affiliation{\CarlPhys}

\author{\\Jay D. Tasson\,\orcidlink{0000-0002-4777-5087}}
\email{jtasson@carleton.edu}
\affiliation{\CarlPhys}

\author{Jolien D. E. Creighton\,\orcidlink{0000-0003-3600-2406}}
\affiliation{\UWM}

\begin{abstract}

The speed of gravitational waves $v_g$ can be measured with the time delay between gravitational-wave detectors. Our study provides a more precise measurement of $v_g$ using gravitational-wave signals only, compared with previous studies.  We select 52 gravitational-wave events that were detected with high confidence by at least two detectors in the first three observing runs (O1, O2, and O3) of Advanced LIGO and Advanced Virgo. We use Markov chain Monte Carlo and nested sampling to estimate the $v_g$ posterior distribution for each of those events. We then combine their posterior distributions to find the 90\% credible interval of the combined $v_g$ distribution for which we obtain $0.99^{+0.02}_{-0.02}c$ without the use of more accurate sky localization from the electromagnetic signal associated with GW170817.  Restricting attention to the 50 binary black hole events generates the same result, while the use of the electromagnetic sky localization for GW170817 gives a tighter constraint of $0.99^{+0.01}_{-0.02}c$.  The abundance of gravitational wave events allows us to apply hierarchical Bayesian inference on the posterior samples to simultaneously constrain all nine coefficients for Lorentz violation in the nondispersive,  nonbirefringent limit of the gravitational sector of the Standard-Model Extension test framework. We compare the hierarchical Bayesian inference method with other methods of combining limits on Lorentz violation in the gravity sector that are found in the literature.
\end{abstract}

\maketitle

\section{Introduction} 
\label{sec:intro}

The third observing run (O3) of Advanced LIGO~\cite{aligo} and Advanced Virgo~\cite{avirgo} was the first complete run in which all three detectors were used \cite{LVCGWTC2, LVCGWTC3}. In total, O3 adds 79 gravitational-wave (GW) candidates, more than seven times the 11 GW candidates from the first (O1) and second (O2) observing runs combined \cite{LVCGWTC1}. With the availability of many more GW events, it becomes possible to measure the speed of GWs $v_g$ more precisely than previous works that used similar methods \cite{Cornish:2017sgw, O12sog}.  Furthermore, it allows a direct and comprehensive exploration of the isotropy of $v_g$ for the first time.

General relativity (GR) predicts that the speed of GWs is the same as the vacuum speed of light $c$. The GWs detected by Advanced LIGO and Advanced Virgo can be used to make statistical inferences about $v_g$, thereby testing the theory of GR. The first measurement of $v_g$ using the time delay between the GW detectors was performed by Ref.~\cite{Cornish:2017sgw}. By applying Bayesian inference, the 90\% credible interval of $v_g$ distribution was constrained to be $(0.55c, 1.42c)$ \cite{Cornish:2017sgw}. Reference \cite{O12sog} further constrained the 90\% credible interval to $(0.97c, 1.05c)$, by applying similar methods to 11 events from O1 and O2. With a total of 52 high-confidence multidetector GW events accrued through the end of O3, we are able to perform a similar analysis using more events, more robustly testing the theory of GR. The method used here remains much less sensitive than the multimessenger astronomy approach used in Ref.~\cite{abbott:2017bnsgrb}, which placed the constraint $+7 \times 10^{-16} \leq \frac{v_g-c}{c}\leq -3 \times 10^{-15}$.  Rather an improved precision, the present approach provides confirmation of the basic conclusion of $v_g=c$ via an alternative approach.  More significantly, the many events available that come from different sky directions permit the exploration of the isotropy of $v_g$. 

The large body of events now available, which arrive from a multitude of sky directions, allows for a complete exploration of the isotropy of $v_g$ in the context of the Lorentz invariance test framework provided by the gravitational Standard-Model Extension (SME).\footnote{For an annually updated review of observational and experimental results, see Ref.~\cite{data}. For early foundational work on the SME, see Ref.~\cite{ck}. For foundational gravity-sector work, see Ref.~\cite{akgrav, lvpn}.} Reference \cite{O12sog} simultaneously constrained the first four of nine coefficients for Lorentz violation in the nondispersive, nonbirefringent limit of the gravity sector using four GW events from O1 and O2. Other recent works \cite{bire1, bire2, disp} have sought the effects of birefringence and dispersion using the SME. 
 Still others have sought the dependence of GW speed on the motion of the source \cite{ghosh2023test}. In this paper, we use 24 of 52 high-significance multidetector GW events to simultaneously constrain all nine coefficients in the nondispersive, nonbirefringent limit of the SME.  While our constraints are much weaker than previous works such as Ref.~\cite{170817mma}, which have constrained the coefficients for Lorentz violation in the gravity sector down to the order of $10^{-15}$ to $10^{-14}$ via multimessenger astronomy, these constraints were obtained using models with only one parameter each. Therefore, our work is the first to provide direct limits from GW observations on all nine coefficients simultaneously.

The remainder of this paper is organized as follows.  In Sec.\ \ref{sec:sog}, we discuss the methods used to extract $v_g$ estimates for each event and present the results.  Section \ref{sec:sme} presents and compares a number of methods for extracting simultaneous limits on the nine coefficients for Lorentz violation before presenting our final estimate of these coefficients from the O1--O3 data.

\section{Speed of Gravitational Waves} 
\label{sec:sog}

\subsection{Bayesian inference methods \label{sec:measureSog}}
Here, we briefly describe our method for obtaining the speed of GWs. Interested readers are invited to refer to Ref.~\cite{O12sog} for full details. 

When a GW passes through Earth, if two or more detectors detect the signal, we can use the relative locations of the detectors and the difference in detection times from those detectors to simultaneously estimate the sky location of the GW event and $v_g$. With only one detector, we cannot find any $v_g$ information, as there is no difference in detection times in this case. Therefore, we select those events that are detected by at least two GW detectors.

Furthermore, we only consider those events whose median signal-to-noise ratios (SNR) are no smaller than 10.0, as reported in the GWTC-2 and GWTC-3 catalog papers \cite{LVCGWTC2, LVCGWTC3}. In total, 41 O3 events meet our selection criteria and are listed in Tables \ref{table:O3a} and \ref{table:O3b}. All O1 and O2 events meet these two selection criteria, so we include their posterior distributions used in Ref.~\cite{O12sog} in our analysis.  Note that the SNR values used to select the O1 and O2 events (which are the same as what are used in Ref.~\cite{O12sog}) correspond to the network SNR with which the events were found by the \textsc{GstLAL} search pipeline as reported in Ref.~\cite{LVCGWTC1}. 

The standard parameter estimation using GW data from multiple detectors imposes the constraint that GWs travel at the speed of light \cite{lalsuite}. In this work, we remove this constraint such that $v_g$ becomes a parameter to be estimated with all other signal parameters. This causes wider distributions for certain parameter estimations. For example, the calculated sky area is often larger because a defined $v_g$ aids sky localization.  

Gravitational wave data $d$, can be decomposed into a pure GW signal $h(t)$ plus random noise $n(t)$,
\begin{equation}
d(t) = h(t) + n(t).
\end{equation} 
Within the framework of Bayesian inference, the posterior distribution of the parameters $\vec{\theta}$ characterizing a GW signal is computed from the likelihood of obtaining GW data given particular values of said parameters and the {\it a priori} knowledge of what we expect those values to be. The likelihood function is constructed by assuming the noise $n(t)$ to be stationary and Gaussian distributed. For details regarding the exact forms of the likelihood see~Ref.~\cite{O12sog}. Once obtained, the joint posterior distribution of the signal parameters can be used to compute the marginalized posterior distribution of $v_{g}$ as in
\begin{equation}
p(v_g|d) = \int p( \vec{\theta}|d)  d\vec{\theta}^\prime,
\end{equation} 
where $\vec{\theta}^\prime$ is the set of parameters in $\vec{\theta}$ except for $v_g$ \cite{O12sog}. 

To carry out parameter estimation for each event that passes our selection criteria, we use public data \cite{gwosco2,gwosco3} from GWTC-1 through GWTC-3. We use \textsc{\texttt{lalinference\_mcmc}} \cite{lalsuite, Metrolis, Hastings, Veitch:2015}, which implements Markov chain Monte Carlo (MCMC) with the Metropolis-Hastings algorithm and \textsc{\texttt{lalinference\_nest}}, which implements nested sampling to run the Bayesian parameter estimation \cite{lalsuite, lalinf_nest1, lalinf_nest2}. For our purposes of extracting $v_g$ distributions, these two algorithms generate comparable results. We use the publicly available power spectral densities and calibration envelopes from the LIGO Scientific, Virgo and KAGRA (LVK) Collaboration in our analysis. In this paper, we use a uniform prior in $v_g$ between $0.1c$ and $10c$. When the $v_g$ posterior rails against the prior, we increase the upper limit of the prior by another $10c$. The broadest prior we use is from $0.1c$ to $30c$, which we only use for one event, GW190929\_012149. For parameters such as binary masses and spins, we use the same uniform and isotropic priors as those used by the LVK \cite{LVCGWTC1, LVCGWTC2, LVCGWTC3}. We choose a distance prior that is proportional to luminosity distance squared, similar to Ref.~\cite{LVCGWTC1}. We do not use the more complicated cosmological priors used by Refs. \cite{LVCGWTC2,LVCGWTC3}.  For O1 and O2 events, we use the posterior samples from Ref.~\cite{O12sog}, which used the \textsc{\texttt{IMRPhenomPv2}} \cite{imrphenomd, imrphenompv2-1, imrphenompv2-2} waveform for all events except for the binary neutron star (BNS) event GW170817, which was analyzed with the \textsc{\texttt{TaylorF2}} waveform \cite{taylorf2-1, taylorf2-2, taylorf2-3, taylorf2-4, taylorf2-5, taylorf2-6}. For most O3 events, we use the \textsc{\texttt{IMRPhenomD}} waveform \cite{imrphenomd, imrphenompv2-1}, which is an aligned spin waveform model for black-hole binaries. We do not use the more sophisticated \textsc{\texttt{IMRPhenomPv2}} model for these events since in the context of our study, we do not expect any significant change in $v_g$ measurements to result from the additional intricacies of the more sophisticated model. We have verified this lack of change for a subset of these events and hence chosen to stick to the \textsc{\texttt{IMRPhenomD}} model consistently for all O3 events except for GW190521. For additional discussion of this point, see Appendix A.  For the extremely high-mass binary black hole (BBH) event GW190521, we use the \textsc{\texttt{NRSur7dq4}} waveform \cite{NRSur7dq4}, which is one of the waveform models used by Ref.~\cite{LVC190521} for inferring this event's source properties. We note that \textsc{\texttt{IMRPhenomPv2}}, \textsc{\texttt{IMRPhenomD}}, and \textsc{\texttt{NRSur7dq4}} are all waveform models with inspiral,  merger, and ringdown.

We can achieve a more precise measurement of $v_{g}$ by combining data from multiple GW events. By interpreting each observation as an independent experiment, we can multiply the marginalized likelihood as a function of $v_g$ corresponding to each event and obtain the joint posterior distribution of $v_g$ given data from multiple events. For a uniform prior on $v_g$, the joint posterior can be expressed as a product of individual event posteriors. 

Suppose the GW detectors observe $n$ independent GW events with data $d_1, d_2, ...,d_n$. For a uniform prior distribution of $v_{g}$, the combined posterior distribution of $v_{g}$ is
\begin{equation}
p(v_g|d_1,d_2,...,d_n) \propto p(v_g|d_1)p(v_g|d_2)\cdots p(v_g|d_n)\label{multiply}.
\end{equation} 

The single event posterior distributions $p(v_g|d_i)$ are obtained as a numerical function of $v_g$ from its parameter estimation samples by means of Gaussian kernel density estimation~(KDE) \cite{Silverman1986,Scott1992}. We use the package \textsc{Scipy}'s implementation of Gaussian KDE to obtained the posteriors~\cite{2020SciPy-NMeth}. The joint posterior distribution is then obtained through Eq.~\eqref{multiply}.

Then, for individual and combined posteriors, we calculate Bayes factors $K$, via the Savage-Dickey density ratio 
\begin{equation}
    K = {p(v_g=c\:|\:d_1,d_2,...)\over p(v_g=c)},
\end{equation}
where $p(v_g=c\:|\:d_1,d_2,...)$ is the posterior probability of $v_g=c$ and $p(v_g=c)$ is the prior probability of $v_g=c$ \cite{bayes-sd}. Higher Bayes factors suggest stronger evidence for $v_g=c$.

\subsection{Results}
In Tables \ref{table:O3a} and \ref{table:O3b}, we show the $v_g$ estimates with 90\% credible intervals, network SNRs, sky areas at 90\% credible level, and Bayes factors for the selected 41 O3 events.  Also shown are the analogous quantities obtained from their combined posteriors. Out of the 41 selected O3 events, 40 events are BBH candidate events. GW200115\_042309 is a neutron star--black hole (NSBH) event, with masses of $5.9^{+2.0}_{-2.5} \mathrm{M}_{\odot}$ and $1.44^{+0.85}_{-0.29} \mathrm{M}_{\odot}$ at 90\% credible interval \cite{LVCGWTC3}. Here, by combining the 41 selected O3 events, we constrain the 90\% credible interval of $v_g$ to be $0.99^{+0.02}_{-0.03}c$, with a Bayes factor of $205.9$. 

We combine the O3 results with the O1 and O2 results discussed in Ref.\ \cite{O12sog}. The 11 O1 and O2 events are run with \textsc{\texttt{lalinference\_mcmc}}, which shows results that are consistent with \textsc{\texttt{lalinference\_nest}} used for O3a runs \cite{O12sog, LVCGWTC1, LVCGWTC2}. In Table \ref{table:O12}, we show the $v_g$ estimates with 90\% credible intervals, network SNRs, sky areas at 90\% credible level, and Bayes factors for the 11 O1 and O2 events and their combined posteriors. We use the same posterior samples as used by Ref.~\cite{O12sog}, but Table \ref{table:O12} shows slightly different 90\% credible intervals from those in Ref.~\cite{O12sog} because we use Gaussian KDE smoothing in this study to extract the credible intervals while Ref.~\cite{O12sog} directly used the posterior samples without KDE smoothing \cite{O12sog}. These 11 events were detected by at least two detectors and had median \textsc{GstLAL} network SNR values greater than 10.0 \cite{LVCGWTC1}. GWTC-2.1 \cite{LVCGWTC21} shows network SNR values for O1 and O2 events based on \textsc{\texttt{lalinference}} parameter estimations, but we choose \textsc{GstLAL} SNR values to be consistent with Ref.~\cite{O12sog} from which we obtain the $v_g$ posterior samples. GW170817 is a BNS event that was also detected in the electromagnetic spectrum \cite{LVC:GW170817, 170817mma}. The ``fixed" label means that the result uses the sky localization from the electromagnetic detections, which is much more precise than the localization generated by GW detection pipelines.

\begin{figure}[h]
\centering
\includegraphics[width=0.5\textwidth]{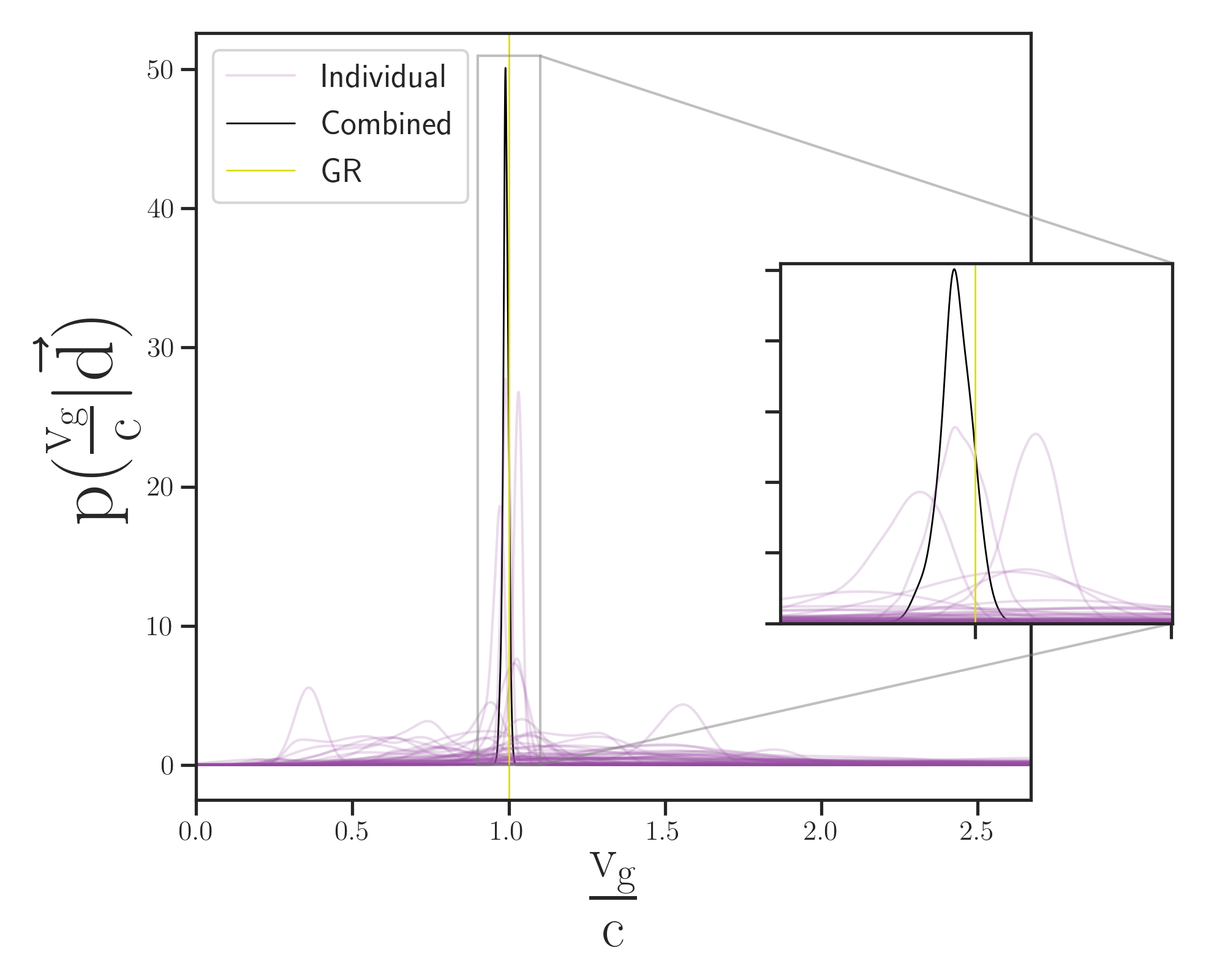}
\caption{\label {fig:combined} Posterior distributions of $v_g$ inferred jointly from all 41 events. The black solid line represents the joint posterior, the thin purple lines represent individual event posteriors and GR ($v_g=c$) is marked by the yellow vertical line.}
\end{figure}

Combining the 41 O3 events and 11 O1 and O2 events without fixing GW170817's sky localization at the detected electromagnetic (EM) signal, we obtain the 90\% credible interval of $v_g$ to be $0.99^{+0.02}_{-0.02}c$, with a Bayes factor of $291.9$. With GW170817 sky localization fixed, $v_g$ is $0.99^{+0.01}_{-0.02}c$, with a Bayes factor of $249.0$. For a total of 49 BBH events, i.e. excluding GW170817, GW190924, and GW200115, $v_g$ is $0.99^{+0.02}_{-0.02}c$, with a Bayes factor of $221.2$. Figure \ref{fig:combined} shows the combined posterior of $v_g$.

\begin{table} [h]
\scriptsize
\centering
\rowcolors{2}{}{lightgray}
\renewcommand{\arraystretch}{2}
\newcolumntype{S}{@{\centering\arraybackslash}m{0.45cm}}
\begin{tabularx}{\linewidth}{>{\centering\arraybackslash}m{3cm}S@{\centering\arraybackslash}m{1.15cm}S@{\centering\arraybackslash}m{0.6cm}S@{\centering\arraybackslash}m{0.82cm}S@{\centering\arraybackslash}m{1.2cm}}
\hline
O3a Event & \hspace{1.25cm}& $v_{g}(\mathrm{c})$ &\hspace{1.25cm} & SNR  &\hspace{1.25cm} & $\Omega$($\mathrm{deg}^2$) &\hspace{1.25cm} & Bayes factor\\
\hline
$^*$GW190408\_181802 & & $1.66^{+0.72}_{-0.89}$ && 15.3 && 1216 && 3.5\\
$^*$GW190412 && $1.49^{+0.47}_{-0.53}$ && 18.9 && 594 && 2.8\\
GW190421\_213856 && $1.15^{+0.46}_{-0.57}$ && 10.7 && 2837 && 11.9\\
$^*$GW190503\_185404 & & $0.55^{+0.26}_{-0.24}$ && 12.4 && 1237 && 0.5\\
$^*$GW190512\_180714 && $1.42^{+0.74}_{-0.98}$ && 12.2 && 1637 && 4.9\\
$^*$GW190513\_205428 && $1.26^{+1.51}_{-0.65}$ && 12.9 && 1075 && 7.6\\
GW190517\_055101 && $0.88^{+0.67}_{-0.38}$ && 10.7 && 2125 && 13.8\\
GW190519\_153544 && $2.04^{+1.75}_{-1.21}$ && 15.6 && 2070 && 3.7\\
GW190521 && $1.82^{+5.01}_{-1.02}$ && 14.2 && 2279 && 3.4\\
GW190521\_074359 && $1.20^{+0.64}_{-0.44}$ && 25.8 && 2318 && 11.8\\
$^*$GW190602\_175927 && $0.98^{+0.17}_{-0.65}$ && 12.8 && 1567 && 27.1\\
GW190630\_185205 && $6.07^{+3.52}_{-4.90}$ && 15.6 && 3983 && 0.5\\
$^*$GW190701\_203306 && $0.85^{+0.21}_{-0.43}$ && 11.3 && 203 && 21.4\\
GW190706\_222641 && $6.67^{+2.91}_{-4.88}$ && 12.6 && 3726 && 0.3\\
GW190707\_093326 && $2.77^{+5.01}_{-1.69}$ && 13.3 && 5708 && 0.9\\
$^*$GW190720\_000836 && $1.54^{+0.13}_{-0.37}$ && 11.0 && 388 && 1.3\\
$^*$GW190727\_060333 && $3.42^{+2.71}_{-2.05}$ && 11.9 && 1521 && 0.6\\
$^*$GW190728\_064510 && $0.97^{+0.82}_{-0.53}$ && 13.0 && 1873 && 14.9\\
$^*$GW190814 && $1.33^{+0.43}_{-0.28}$ && 24.9 && 334 && 4.1\\
GW190828\_063405 && $6.11^{+3.41}_{-4.90}$ && 16.2 && 2498 && 0.5\\
GW190828\_065509 && $1.64^{+4.77}_{-0.99}$ && 10.0 && 2917 && 4.0\\
$^*$GW190915\_235702 && $0.40^{+0.71}_{-0.11}$ && 13.6 && 904 && 5.1\\
$^*$GW190924\_021846 && $0.92^{+0.32}_{-0.32}$ && 11.5 && 918 && 21.1\\
GW190929\_012149 && $6.08^{+3.53}_{-4.90}$ && 10.1 && 5761 && 0.5\\
\hline
\end{tabularx}
\caption{The 90\% credible intervals of $v_g$ from individual O3a events posteriors. Median network SNR values are reported from GWTC-2 \cite{LVCGWTC2}. The 90\% credible regions of the sky localization ($\Omega$) without fixing $v_g$ at $c$ are calculated from the individual posteriors \cite{LVCGWTC2}. The Bayes factor $K$ indicates how strong the posterior distributions support $v_g=c$. The asterisks ($^*$) in front of GW event names represent the GW events chosen for obtaining constraints on all nine coefficients for Lorentz violation in Sec. \ref{sec:sme}.\\}
\label{table:O3a}
\end{table}

\begin{table} [h]
\scriptsize
\centering
\rowcolors{2}{}{lightgray}
\renewcommand{\arraystretch}{2}
\newcolumntype{S}{@{\centering\arraybackslash}m{0.45cm}}
\begin{tabularx}{\linewidth}{>{\centering\arraybackslash}m{3cm}S@{\centering\arraybackslash}m{1.15cm}S@{\centering\arraybackslash}m{0.6cm}S@{\centering\arraybackslash}m{0.82cm}S@{\centering\arraybackslash}m{1.2cm}}
\hline
O3b Event & \hspace{1.25cm}& $v_{g}(\mathrm{c})$ &\hspace{1.25cm} & SNR  &\hspace{1.25cm} & $\Omega$($\mathrm{deg}^2$) &\hspace{1.25cm} & Bayes factor\\
\hline
GW191109\_010717 && $1.80^{+1.20}_{-0.93}$ && 17.3 && 4033 && 2.4\\
GW191129\_134029 && $1.69^{+4.11}_{-1.18}$ && 13.1 && 4891 && 4.2\\
GW191204\_171526 && $1.18^{+1.26}_{-0.99}$ && 17.5 && 3009 && 5.5\\
GW191215\_223052 && $1.44^{+2.02}_{-0.82}$ && 11.2 && 3280 && 5.9\\
GW191216\_213338 && $1.31^{+0.59}_{-0.37}$ && 18.6 && 2076 && 11.0\\
GW191222\_033537 && $5.11^{+4.19}_{-3.66}$ && 12.5 && 3206 && 0.4\\
GW191230\_180458 && $1.42^{+1.07}_{-0.91}$ && 10.4 && 2538 && 5.9\\
GW200115\_042309 && $3.02^{+5.76}_{-2.21}$ && 11.3 && 3271 && 1.2\\
GW200128\_022011 && $5.57^{+3.79}_{-3.90}$ && 10.6 && 8988 && 0.3\\
$^*$GW200129\_065458 && $0.99^{+0.08}_{-0.38}$ && 26.8 && 149 && 53.7\\
$^*$GW200202\_154313 &&  $0.69^{+0.20}_{-0.33}$ && 10.8 && 1551 && 0.7 \\
$^*$GW200208\_130117 && $1.39^{+0.40}_{-0.60}$ && 10.8  && 706 && 5.6\\
GW200219\_094415 && $1.92^{+2.20}_{-1.59}$ && 10.7 && 3114 && 3.2 \\
$^*$GW200224\_222234 && $1.03^{+0.02}_{-0.04}$ && 20.0 && 94 && 48.1 \\
GW200225\_060421 && $1.24^{+0.66}_{-0.75}$ && 12.5 && 3583 && 8.0 \\
$^*$GW200311\_115853 && $0.96^{+0.03}_{-0.05}$ && 17.8 && 102 && 32.1\\
$^*$GW200316\_215756 && $3.76^{+5.26}_{-2.76}$ && 10.3 && 1881 && 0.8 \\
\hline
O3 combined (BBHs) && $0.99^{+0.02}_{-0.02}$ && && && 203.3\\[1.5ex]
O3 combined && $0.99^{+0.02}_{-0.03}$  && && && 205.9\\[1.5ex]
\hline
\end{tabularx}
\caption{The 90\% credible intervals of $v_g$ from individual O3b events posteriors and combined posteriors using all O3 events. Median network SNR values are reported from GWTC-3 \cite{LVCGWTC3}. The 90\% credible regions of the sky localization ($\Omega$) without fixing $v_g$ at $c$ are calculated from the individual posteriors \cite{LVCGWTC3}. The Bayes factor $K$ indicates how strong the posterior distributions support $v_g=c$. The asterisks ($^*$) in front of GW event names represent the GW events chosen for obtaining constraints on all nine coefficients for Lorentz violation in Sec. \ref{sec:sme}.\\}
\label{table:O3b}
\end{table}

\begin{table} [h]
\scriptsize
\centering
\rowcolors{2}{}{lightgray}
\renewcommand{\arraystretch}{2}
\newcolumntype{S}{@{\centering\arraybackslash}m{0.375cm}}
\begin{tabularx}{\linewidth}{>{\centering\arraybackslash}m{3.3cm}S@{\centering\arraybackslash}m{1.15cm}S@{\centering\arraybackslash}m{0.6cm}S@{\centering\arraybackslash}m{0.82cm}S@{\centering\arraybackslash}m{1.2cm}}
\hline
Event & \hspace{1.25cm}& $v_{g}(\mathrm{c})$ &\hspace{1.25cm} & SNR  &\hspace{1.25cm} & $\Omega$($\mathrm{deg}^2$) &\hspace{1.25cm} & Bayes factor\\
\hline
GW150914 && $0.68^{+0.50}_{-0.29}$ && 24.4 && 2385 && 7.0\\
GW151012 && $6.01^{+3.45}_{-4.41}$ && 10.0 && 6607 && 0.4\\
GW151226 && $4.04^{+5.17}_{-3.28}$ && 13.1 && 6515 && 1.3\\
GW170104 && $1.61^{+1.88}_{-1.12}$ && 13.0 && 5313 && 4.3\\
$^*$GW170608 && $1.14^{+0.22}_{-0.26}$ && 14.9 && 1269 && 16.9\\
$^*$GW170729 && $3.70^{+1.89}_{-2.12}$ && 10.8 && 1287 && 0.3\\
GW170809 && $0.65^{+0.36}_{-0.34}$ && 12.4 && 2252 && 8.6\\
$^*$GW170814 && $1.00^{+0.09}_{-0.28}$ && 15.9 && 250 && 68.0\\
GW170817 (unfixed) && $1.01^{+0.04}_{-0.06}$ && 33.0 && 53 && 111.9\\
$^*$GW170817(fixed) && $0.99^{+0.03}_{-0.02}$ && 33.0 && 0 && 228.5\\
$^*$GW170818 && $0.94^{+0.24}_{-0.38}$ && 11.3 && 168 && 20.5\\
GW170823 && $3.90^{+4.95}_{-3.00}$ && 11.5 && 6412 && 0.9\\
\hline
Combined (All BBHs) && $0.99^{+0.02}_{-0.02}$ && && && 221.2\\[1.5ex]
Combined (All, fixed) && $0.99^{+0.01}_{-0.02}$ && && && 249.0\\[1.5ex]
Combined (All, unfixed) && $0.99^{+0.02}_{-0.02}$  && && && 291.9\\[1.5ex]
\hline
Combined (O1/2, fixed) && $0.99^{+0.02}_{-0.02}$ && && && 249.0\\[1.5ex]
Combined (O1/2, unfixed) && $1.01^{+0.04}_{-0.05}$ && && && 149.0\\[1.5ex]
\hline
\end{tabularx}
\caption{The 90\% credible intervals of $v_g$ from individual O1 and O2 events posteriors. Combined posteriors are presented for all selected events (O1, O2, and O3) as well as for O1 and O2 events. Network SNR values are reported from the \textsc{GstLAL} search pipeline in GWTC-1 \cite{O12sog, LVCGWTC1}. The 90\% credible regions of the sky localization ($\Omega$) without fixing $v_g$ at $c$ are calculated from the individual posteriors \cite{O12sog, LVCGWTC1}. The Bayes factor indicates how strongly the posterior distributions support $v_g=c$. The asterisks ($^*$) in front of GW event names represent the GW events chosen for obtaining constraints on all nine coefficients for Lorentz violation in Sec. \ref{sec:sme}. The ``fixed" and ``unfixed" labels represent whether we fixed the sky localization of GW170817 at the source of its EM counterpart.}
\label{table:O12}
\end{table}

\subsection{Discussion}
In Ref.~\cite{O12sog}, with 11 O1 and O2 events and GW170817's sky localization unfixed, the combined posterior distribution of $v_g$ was measured to be $1.01^{+0.04}_{-0.05}c$, while here we measure $v_g$ to be $0.99^{+0.02}_{-0.02}c$ with the 52 selected events. With GW170817's localization fixed, in Ref.~\cite{O12sog}, the combined posterior distribution of $v_g$ was measured to be $0.99^{+0.02}_{-0.02}c$ for 11 events, while here we find $0.99^{+0.02}_{-0.01}c$ for the 52 events. Given that $1c$ is the GR prediction for $v_g$, the combined posterior distributions of $v_g$ measured using 52 selected events show no evidence for a violation of GR. All of these combined results have Bayes factors on the order of $10^2$, providing strong evidence for $v_g=c$.

Here, our measured distribution of $v_g$ is much narrower than that measured with 11 O1 and O2 events in Ref.~\cite{O12sog} using GW signals alone. This is reasonable, given the larger sample size of events included in this study. When we assume that the $v_g$ distributions of individual events are independent and identically distributed, we expect that the measurement errors would decrease by $1/\sqrt{n}$. In our calculations, we find that the combined $v_g$ distribution roughly follows such a pattern as more events are added. For example, with 11 O1 and O2 events, the combined $v_g$ posterior had an error bar of $0.09c$. With 52 events in total, the combined posterior had an error bar of $0.04c$, which follows $0.09c/\sqrt{52/11}\approx0.04c$. With GW170817's sky localization unfixed, we find that Bayes factor more than doubles from the value of $149.0$ obtained from 11 O1 and O2 events to the value of $291.9$ obtained with all 52 events. This, in conjunction with the error bar being reduced by half, implies that our measurement with 52 GW events in total has provided approximately twice stronger evidence for $v_g=c$.

Interestingly, we find that the combined 90\% $v_g$ credible interval using the 41 O3 events is approximately the same as the 90\% $v_g$ credible interval obtained by only considering GW170817 with the fixed sky localization. GW170817 had an SNR of $33.0$, while only four of the 41 O3 events had SNRs above $20.0$, with the highest being $26.8$ for GW200129\_065458. The similarity between the $v_g$ posterior of GW170817 alone and the 41 O3 events combined suggests that some combination of higher SNRs and better sky localization do help put tighter constraints on $v_g$. This shows that our decision to exclude events with SNRs lower than $10.0$ should not have a high impact on the $v_g$ estimates.

Looking to the future, additional two- and three-detector BBH events with SNRs typical of those above will lead to a slow improvement in $v_g$ measurements as improvements proceed as $1/\sqrt{n}$.  However, as GW detectors become more sensitive and the network of detectors expands, we expect more high-SNR, multidetector GW events that would likely lead to a more rapid pace of progress in $v_g$ estimations via the methods used here. Meanwhile, future multimessenger detections can provide more precise sky localizations, which will likely improve the error bars on the 90\% $v_g$ credible interval further.

\section{Simultaneous SME Limits} 
\label{sec:sme}

\subsection{Basics}
In the nonbirefringent, nondispersive limit of the SME (mass dimension $d=4$), using natural units and assuming that the nongravitational sectors, including the photon sector, are Lorentz invariant, the difference between the group velocities of gravity and light takes the form
\cite{kmgw} \hfill
\begin{equation}
   \Delta v = -\sum_{lm} Y_{lm}(\hat{n}) \frac{1}{2} (-1)^{1+l}\bar{s}_{lm},
   \label{master1}
\end{equation}
where the $Y_{lm}$'s are the spherical harmonics with $l\leq~2$.  Here the nine Lorentz-violating degrees of freedom are characterized by the spherical coefficients for Lorentz violation $\bar{s}_{lm}$, and $\hat{n}$ is the sky location of the source of the GWs. We can expand Eq.(\ref{master1}) over positive $m$ to get its equivalent expression: \bea
\nonumber
\Delta v
=
\sum_l (-1)^l \Big( \half \sb_{l0} Y_{l0} 
  + \sum_{m>0} && [\re \sb_{lm} \re Y_{lm}\\
  &&-  \im \sb_{lm} \im Y_{lm}]\Big). \phantom{222}
\label{eq:master2}
\eea

The SME is a broad and general test framework for testing Lorentz invariance.  Unlike models that attempt to describe specific effects with a small number of parameters, test frameworks, because of their generality, have a large number of undetermined coefficients to be explored in experimental data.  While a number of studies have proceeded under a simplified approach, sometimes referred to as a maximum reach analysis \cite{flowers2016}, in which only one coefficient at a time is considered, it is also common to study a family of coefficients together in what is sometimes referred to as a coefficient separation approach \cite{flowers2016}.  In the context of the maximum reach approach, many coefficients can sometimes be constrained one at a time using a single measurement, while a number of measurements that is greater than or equal to the number of coefficients considered is typically required to simultaneously measure the entire family.

A number of approaches to simultaneously estimate multiple coefficients exist in the literature \cite{microscope,cerenkov,O12sog,bire1}.  One approach involves directly fitting a single data stream to a model involving all of the coefficients in the family.\footnote{For a recent example of this approach in the gravity sector, see Ref.~\cite{microscope}.} This approach is well suited to experiments that take data as the laboratory is boosted and rotated.  

In the context of astrophysical observations, each individual event provides a measurement of a linear combination of coefficients for Lorentz violation.  A system of these inequalities must then be solved, or otherwise disentangled, for estimates of the coefficients for Lorentz violation.  Several methods of addressing this issue exist in the literature.  In this section, we will compare the implications of several of these approaches in the context of the speed of GW data, as well as introduce new methods based on hierarchical Bayesian inference.  Our goal is to consolidate information about these methods and help illuminate their relative merits. We achieve that goal by performing a mock data challenge (MDC), where we generate synthetic data corresponding to a chosen set of ``true" values of the SME coefficients and test the efficacy of each method in recovering the true values from the synthetic data. 

Given their performance in the MDC, along with other considerations, we choose one of these methods whose merits outweigh that of the others and use it to analyze the real speed of gravity data from a subset of the events analyzed in Sec.~\ref{sec:sog} to generate the final results of our SME analysis.  Because well-localized events are most informative for the SME analysis, we choose the 24-event subset of those considered in Sec.~\ref{sec:sog} with 90\% credible posterior sky areas under 2000 deg$^2$ as obtained from our parameter estimation with $v_g$ as a free parameter. 

\subsection{Linear programming method}
A number of past studies that have performed an analysis using limits from astrophysical events have taken a linear programming approach.  See, for example, Refs.\ \cite{matt, mike, cerenkov}.  The basic idea translated to the speed of GWs problem proceeds as follows.  

From a given event we have an upper and lower bound on $\Delta v$.  If we suppose that we know an exact sky location, as is effectively the case for GW170817 when the electromagnetic signal's localization is used, then Eq.\ \rf{eq:master2} can be understood as generating a pair of hyperplanes in $\sb_{lm}$ space that are the boundaries of the parameter space excluded by the event.  A subsequent event at a different sky location will generate a distinct pair of hyperplanes.  Once a set of $n$ events is collected at distinct sky locations, where $n$ is greater than or equal to the dimensionality of the coefficient space, then a finite maximum and minimum allowed value for each coefficient can be identified via a linear programming scheme such as the simplex method. 

In the applications of Refs. \cite{matt, mike, cerenkov}, the sky localizations were sufficiently well known that analysis could proceed directly via the above prescription.  In the current problem, for all events except GW170817, the sky localization is comparatively poorly known.  This makes the slopes of the hyperplanes bounding the allowed region poorly known.  

To address our uncertainty in sky positions, the linear programming scheme can be adapted as follows.  The linear programming process can be applied with all possible hyperplanes generated by samples from our inference that fall within the 68\% credible sky localization bands.  The worst-case limits generated by the set of linear programming analysis can then be taken as bounds.  As might be expected, this method generates very conservative bounds relative to the methods to follow.  Testing this approach using four test events and a sky map resolution of $N_{\rm side}=64$, which corresponds to $12\times 64^2=49152$ pixels on the celestial sphere \cite{healpy1, healpy2}, we generate bounds that are about an order of magnitude greater than the $1\sigma$ credible intervals found via the application of the random draw method that we present in the next subsection.  Hence, we do not consider this approach further as a method of extracting SME limits from the speed of GWs data at this time.

\subsection{Random draw method}
In Ref.\ \cite{O12sog}, the random draw method for extracting simultaneous limits on coefficients for Lorentz violation was first used.  In that work, simultaneous limits were achieved for the set of four  $\sb_{lm}$ coefficients with $l\leq 1$ by using the four high-confidence, well-localized events available at the time.  In this section, we review this method and discuss ways of extending it to cases in which the number of events exceeds the number of coefficients to be estimated.

The result of the inference discussed in Sec.\ \ref{sec:measureSog} is a set of samples with each sample consisting of values for each of the parameters including the speed of GWs and the sky localization.  Hence distributions for each of the sampled parameters are generated.  If one randomly draws one sample associated with each event, one can then solve for the coefficients for Lorentz violation that are consistent with that set of samples using Eq.\ \rf{eq:master2}.  The process of randomly drawing one sample from each event and solving for the coefficients can be iterated to build up a set of samples for the $\sb_{lm}$ coefficients.  In other words, a set of points in $\sb_{lm}$ space is built up.  

The process described above is straightforward when the number of events observed is equal to the number of coefficients for Lorentz violation to be estimated. Furthermore, in such a scenario, quantile ranges of $\sb_{lm}$ computed from the set of samples of $\sb_{lm}$, accurately represent the uncertainty in our measurement of the SME coefficients. This is because using one posterior sample of $(\Delta v,\theta,\phi)$ from each event and exactly calculating $\sb_{lm}$ from them by solving a set of nondegenerate linear equations, is equivalent to computing and multiplying the posterior distributions of $\sb_{lm}$ for each event and then drawing one sample from that joint posterior. However, in the case where the number of GW observations exceeds the number of SME coefficients, the linear equations become degenerate and hence no longer exactly solvable. While one can be tempted to cherry-pick the top nine events with the highest SNRs and lowest sky areas from the set of observations and perform random draw on those, such an analysis will not be maximally informative given the data we have. We can do better using Bayesian hierarchical inference techniques, which can combine information from a large number of events, producing much more informative bounds on the SME coefficients with accurate estimation of measurement uncertainties. 

Before discussing our robust Bayesian methods we show how the random draw method can be extended to the case in which the number of observations exceeds the number of coefficients for Lorentz violation by means of singular value decomposition (SVD). However, this extension of the random draw method is susceptible to the limitations of the approximation used to perform the SVD and hence cannot produce reliable uncertainty estimates for the measured Lorentz violation parameters. We elaborate more on this near the end of this section while informing the reader beforehand that this SVD-assisted random draw generalization is useful in the present context only as a consistency check and an optimization tool for the hierarchical Bayesian methods on which we rely for our final results.

With $n_{\rm SME}$ coefficients for Lorentz violation  and $N_{E}$ events, with $n_{\rm SME}<N_{E}$, for each random draw, we need to solve the degenerate system of linear equations:
\begin{eqnarray}
    \boldsymbol{A}[\boldsymbol{\bar{s}_{lm}}]&=&[\boldsymbol{\Delta v}].\label{equation:rd1}
\end{eqnarray}
Here $\boldsymbol{A}$ is an $N_E\times n_{\rm SME}$ matrix in which each row corresponds to one of the $N_{E}$ events under consideration.  The entries in each of the $n_{\rm SME}$ columns moving across a given row consist of the coefficients of $\boldsymbol{\bar{s}_{lm}}$ in Eq.\ \eqref{eq:master2}, computed for a random sample of $\theta,\phi$ drawn from the event corresponding to that row.  The $n_{\rm SME}$ SME coefficients to be computed are organized into a column vector denoted $[ \boldsymbol{\bar{s}_{lm}}]$, while $[\boldsymbol{\Delta v}]$ denotes a column vector of the randomly drawn $\Delta v$ corresponding to the samples used in constructing the rows of $\boldsymbol{A}$. Before factorizing the nonsquare matrix, we scale both sides of each line of Eq.\ \eqref{equation:rd1} by the standard deviation of the $\Delta v$ samples corresponding to that event. We define $[\boldsymbol{\sigma_{\Delta v}}]$ to be a column vector in which each element corresponds to the standard deviation of the $\Delta v$ samples from that particular event; then, we can write the scaled version of Eq.\ \eqref{equation:rd1} as
\begin{eqnarray}
    \boldsymbol{A'}[\boldsymbol{\bar{s}_{lm}}]&=&[\boldsymbol{\Delta v'}],
\end{eqnarray}
where 
\begin{eqnarray}
 \boldsymbol{A'}_{ij}&=&\frac{\boldsymbol{A}_{ij}}{[\boldsymbol{\sigma_{\Delta v}}]_i}\\
{[\boldsymbol{\Delta v'}]}_i &=& \frac{[\boldsymbol{\Delta v}]_i}{[\boldsymbol{\sigma_{\Delta v}}]_i}.
\end{eqnarray}
The SVD factorizes the nonsquare matrix $\boldsymbol{A'}$ into two orthogonal square matrices
$\boldsymbol{U}$ and $\boldsymbol{V}$, that are $N_E\times N_E$ and $n_{\rm SME}\times n_{\rm SME}$ respectively, and a diagonal $N_E\times n_{\rm SME}$ matrix $\boldsymbol{\Sigma}$ with nonnegative entries:
\begin{eqnarray}
    \boldsymbol{A'}&=&\boldsymbol{U}\boldsymbol{\Sigma}\boldsymbol{V}^T,
\end{eqnarray}
where $\Sigma$ has the form
\begin{equation}
    \boldsymbol{\Sigma}=\left(\begin{array}{cc}
        \boldsymbol{S} & \boldsymbol{0} \\
        \boldsymbol{0} & \boldsymbol{0}
    \end{array} \right)
\end{equation}
with
\begin{equation}
    \boldsymbol{S}= diagonal\{\sigma_1,...,\sigma_{n_{\rm SME}}\}.
\end{equation}
The nonnegative values $\sigma_1>\sigma_2>...>\sigma_{n_{\rm SME}}$ are known as singular values and are estimated along with $\boldsymbol{U}$ and $\boldsymbol{V}$ by a linear least squares algorithm \cite{Golub1970}. The scaling with the standard deviation of $\Delta v$ essentially transforms a least-square minimized SVD on $\boldsymbol{A}$ into a Chi-square minimized SVD on $\boldsymbol{A'}$. This allows us to properly account for the fact that some events in our list are less significant than others.  Proceeding without this scaling biases the SVD. Once computed, the singular values can be used to solve for $\boldsymbol{\bar{s}_{lm}}$ in Eq.\ \eqref{equation:rd1} :
\begin{equation}
    [\boldsymbol{\bar{s}_{lm}}]_i=\frac{1}{\sigma_i}\sum_{k=1}^{n_{\rm SME}}\boldsymbol{V}_{ik}[\boldsymbol{U}^T\boldsymbol{\Delta v'}]_{k}
\end{equation}
for each draw. We can then estimate the densities of the SME parameters from all draws and produce constraints on them.

We note that despite being a computationally cheap method for computing constraints on the SME coefficients from multiple GW events, the SVD-assisted random draw method has certain inadequacies. There is ambiguity in the exact meaning and interpretation of the uncertainty estimates produced by this method. In the case where the number of events is larger than the number of SME coefficients, this implementation of the random draw method boils down to randomly choosing a posterior sample of $(\Delta v,\theta,\phi)$ from each event and doing a least chi-square fit for the SME parameters. This procedure is then repeated a large number of times, producing a least chi-square fit of the SME coefficients for each draw. However, this is not equivalent to the multiplication of posterior probabilities of the SME coefficients, over all events, and drawing samples from that joint posterior. Thus, the quantile ranges of the set of chi-square fitted SME coefficients do not hold the same meaning as Bayesian credible intervals. While the Bayesian intervals represent regions of the SME parameter space wherein their true values lie with a particular posterior probability given the data, the SVD-based random draw constraints can be expected to have a  different meaning, the exact nature of which remains ambiguous.

Because of these considerations, we conclude that the weighted SVD-assisted random draw method produces constraints that are unreliable and are likely to be underestimates of the true uncertainties in the measurement of SME coefficients. We verify this claim by testing this method against its Bayesian counterparts in a MDC that we describe later in this work. The results of the MDC show that the samples of SME parameters produced by this method are concentrated in a narrow region around the true values of the parameters, which also coincide with the peaks of the posterior distributions inferred by the Bayesian methods. Therein lies the merit of this method in the present context and its potential to serve as a rapid consistency check for the Bayesian methods. Furthermore, this method is extremely fast and computationally cheap and hence can be used to quickly find the narrow region in the parameter space inside which the peak of the posteriors lies. The stochastic MCMC sampling employed by our Bayesian methods is expected to converge much faster if the MCMC chains are initialized near the maxima of the posterior being sampled. Thus the SVD-assisted random draw method can be used to optimize the MCMC sampling used in our Bayesian methods with significant speed-up gains for narrowly peaked SME posteriors. Given the large number of events expected to be observed in O4 and the width of the Bayesian intervals we compute using our current set of events, the posterior distributions of the SME coefficients can be expected to be very narrow post O4, and hence lead to a drastic increase in the computational cost and latency of the Bayesian methods being applied to such a dataset. This will likely make the optimization of the Bayesian methods as offered by the SVD-assisted random draw method a necessary tool in the near future.

\subsection{Hierarchical Bayesian inference}
Since the SME coefficients are properties that are expected to be the same for all events, one can perform Bayesian hierarchical inference on them from the GW data of multiple events. To do so, we can construct the marginalized likelihood of GW data given a particular value of the SME coefficients, jointly from multiple events
\begin{equation}
    \begin{split}
        L(&\bar{s}_{lm}) = \\&\prod_{i\in \{\mathrm{events}\}}\int L(d_i|\Delta v',\theta,\phi)\Pi(\Delta v',\theta, \phi|\bar{s}_{lm})d \Delta v' d\theta d\phi,
        \label{BH1}
    \end{split}
\end{equation}
where the SME sensitive part of the prior imposes the relationship \eqref{eq:master2}  on $\Delta v,\theta,\phi$ for a given value of the SME coefficients :
\begin{equation}
    \begin{split}
        \Pi(\Delta v',\theta, \phi|\bar{s}_{lm}) =\delta(\Delta v'-\Delta v(\bar{s}_{lm},\theta,\phi))\pi(\theta)\pi(\phi).
        \label{BH2}
    \end{split}
\end{equation}
Here, $\Delta v(\bar{s}_{lm},\theta,\phi)$ is the right-hand side of Eq.\ \eqref{eq:master2}. Note that we have chosen to represent the deviation of the speed of gravity from the speed of light by the dummy variable $\Delta v'$ whenever a probabilistic quantity (such as likelihood, posterior, prior, or detection fraction) is expressed as a function of it, so as to distinguish it from the quantity $\Delta v(\bar{s}_{lm},\theta,\phi)$.
The presence of the delta function in Eq.\ \eqref{BH2} is due to the deterministic nature of Eq.~\eqref{eq:master2}. 

By Bayes' theorem, for a uniform prior on $\bar{s}_{lm}$, the likelihood $L(\bar{s}_{lm})$ is proportional to the posterior of these parameters given GW data.  We can now sample this posterior using MCMC to produce joint SME constraints from multiple GW observations.
However, this procedure involves a very large number of evaluations of the likelihoods $L(d_i|\Delta v',\theta,\phi)$, which is so computationally expensive that it is practically infeasible.

To get around this problem, one can again use Bayes' theorem to write the likelihood $L(d_i| \Delta v',\theta,\phi)$ as proportional to the ratio of the posterior $p(\Delta v',\theta,\phi|d_i)$ to the prior:
\begin{eqnarray}
    L(d_i|\Delta v',\theta,\phi) &\propto& \fr{p(\Delta v',\theta,\phi|d_i) }{\pi(\Delta v')\pi(\theta)\pi(\phi)}\label{Bayes}
\end{eqnarray}
Substituting this into Eq.\ \eqref{BH1} gives us
\begin{equation}
    \begin{split}
        L(\bar{s}_{lm}) &\propto\\  \prod_{i\in \{\mathrm{events}\}} &\int p(\Delta v',\theta,\phi|d_i)\delta(\Delta v'-\Delta v(\bar{s}_{lm},\theta,\phi))  \\&d\Delta v' d\theta d\phi. \label{BH3}
    \end{split}
\end{equation}
We can now use the samples drawn from the posterior $p(\Delta v',\theta,\phi|d_i)$ obtained using the parameter estimation run described above to evaluate the integral in Eq.\ \eqref{BH3}. Note that we have ignored a factor of $1/\pi(\Delta v')$ in Eq.\ \eqref{BH3} which is constant since we choose $\pi(\Delta v')$ to be uniform in our parameter estimation runs. However, the presence of the Dirac delta makes it slightly complicated to evaluate this integral directly as a sum over posterior samples. We describe shortly two approximation schemes that can be used to smooth out the discrete sum of Dirac deltas over posterior samples that would entail the evaluation of the integral in Eq.\ \eqref{BH3} and hence constrain the SME coefficients jointly from multiple GW observations. Before that, we first describe why Bayesian inference of this form is subject to selection biases and how we account for them.

Bayesian hierarchical inference from a set of GW events selected based on a particular criterion introduces selection biases into the inferred posterior distribution of hyperparameters \cite{mandel2019,vitale2020}. Since we are selecting events based on whether they were found with a SNR greater than some threshold in at least three detectors, and since each detector has an antenna pattern that makes it more sensitive to certain sky directions than others at the time of detection\cite{Payne2020}, our analysis might be biased toward some values $s_{lm}$ against others. Particularly, the fact that GW search pipelines such as \textsc{GstLAL} only report multidetector coincidences based on whether or not the time delays between the detectors being triggered are smaller than the light travel time between detectors plus a 5 ms window, has the potential to bias our results greatly \cite{Messic2017}. Furthermore, noncoincident events are down-ranked in significance \cite{Messic2017}, making events even less likely to be detectable for certain cases. Other pipelines such as \textsc{PyCBC} use similar methods for identifying multidetector coincidences albeit with a different value for the timing error window~(which is 2 ms for \textsc{PyCBC} \cite{Davies2020}). The existence of this restriction for coincidence formation in search pipelines implies that we are more likely to discover a multidetector event if the speed of GWs is greater than or equal to $c$, as compared to if it were lower than $c$. Thus, our speed of GW measurements may be biased toward measuring $\Delta v\geq 0$ against $\Delta v<0$ along any particular sky position.

 To account for this bias, we must normalize our hierarchical likelihood over the true rate of events as opposed to the detected rate, with the latter being different from the former, due to selection biases. The constant of normalization is the fraction of events that are detectable given a particular value of the hyperparameters and the detection criteria,
 \begin{eqnarray}
    \begin{split}
     L(\bar{s}_{lm}) &\propto \frac{1}{\beta^{N}_{\mathrm{det}}(\bar{s}_{lm})}\\
     \times\prod^N_{i\in \{\mathrm{events}\}} &\int p(\Delta v',\theta,\phi|d_i)\delta(\Delta v'-\Delta v(\bar{s}_{lm},\theta,\phi))  \\&d\Delta v' d\theta d\phi, \label{BHsel1}
     \end{split}
 \end{eqnarray}
 where $\beta_{\mathrm{det}}(\bar{s}_{lm})=\frac{R_{\mathrm{det}}(\bar{s}_{lm})}{R_{\rm true}}$, the fraction of detectable events, is the ratio of the detectable rate of events to the true rate of events \cite{Farr2019}. To calculate the fraction accurately, we must simulate a large number of events whose parameters are drawn from broad enough distributions, inject them into the detector noise realizations, and see what fraction of them are recovered given our selection criteria. To do that, we must first quantify our selection criteria in terms of the parameters that characterize the GW signal. Accurate modeling would require us to recalculate the search pipeline's ranking statistic of a simulated event while allowing for nonzero $\Delta v$ and to find the corresponding false alarm rate (FAR) of that trigger from said ranking statistics. One can then apply a threshold on the combined FAR of the event to classify them as detectable or nondetectable. However, such a calculation would require a pipeline-specific analysis, which is beyond the scope of this work. Instead, we use an approximated selection criteria: for the $i$th event to be detectable, its recovered parameters must satisfy:
 \begin{eqnarray}
    \begin{split}
    {\mathrm{det}} &\implies \{\rho_{H}\geq \rho_{\mathrm{th}},\rho_{L}\geq \rho_{\mathrm{th}},\rho_{V}\geq \rho_{\mathrm{th}}, \rho_{\mathrm{net}}\geq \rho_{\mathrm{net,th}},\\
    & \Delta t_{HL}(\Delta v)\leq\Delta t_{HL}(0)+5\,\mathrm{ms},t_{HV}(\Delta v)\leq\\
    & \Delta t_{HV}(0) +5\,\mathrm{ms}, t_{VL}(\Delta v)\leq \Delta t_{VL}(0)+5\,\mathrm{ms}\},
    \end{split}
 \end{eqnarray}
 where $\rho_{A}$ is the SNR in detector $A$, $\rho_{\rm net}$ is the network SNR, $\Delta t_{AB}(\Delta v)$ is the timedelay of signal arrival between detectors $A$ and $B$ as a function of $\Delta v$, and $\rho_{\rm th}$ is the SNR threshold used for selecting events. Even though we do not select events depending on which search pipeline found them, we use \textsc{GstLAL}'s timing error window to quantify our selection criteria, instead of, say, \textsc{PyCBC}'s, due to the following reason. Among the events that survive our three detector SNR thresholds, most are found by both \textsc{GstLAL} and \textsc{PyCBC} except for GW170818, GW190701, and GW190814, which are found only by \textsc{GstLAL}. Hence, it is sufficient to model the selection biases that might have appeared in this particular study based on \textsc{GstLAL}'s value of the timing error window. This would not have been possible if there were events found by \textsc{PyCBC} and not \textsc{GstLAL} with SNR greater than 10 in three detectors during O3. In such a scenario, a more generalized treatment of selection biases would have been necessary, one that accounts for the difference in timing errors allowed by \textsc{GstLAL} and \textsc{PyCBC}.
 
 Now that we have a quantifiable detection criterion, we can carry out our simulations. Once the simulated events are injected into detector noise realizations and classified as detectable or nondetectable depending on their recovered parameters, it is possible to compute the fraction of events detectable given a choice of CBC parameters:
 \begin{eqnarray}
    f_{\mathrm{det}}(\Delta v',\theta,\phi,\vec{\gamma}) &\propto&\fr{p(\Delta v',\theta,\phi,\vec{\gamma}|det)}{\Pi_{\mathrm{sim}}(\Delta v',\theta,\phi,\vec{\gamma})}\label{BHpdet1}.
 \end{eqnarray}
 Here, $\vec{\gamma}$ are additional CBC parameters such as masses, spins, etc.\ that characterize the waveform, $p(\Delta v',\theta,\phi,\vec{\gamma}|det)$ is the probability of detection, which can be calculated from the set of simulated events that are detectable, and $\Pi_{\mathrm{sim}}$ is the prior from which the simulations are drawn, which has to be broad enough so that we have enough events in both the detectable and nondetectable parts of the parameter space. We can marginalize Eq.\ \eqref{BHpdet1} over suitable priors to get
\begin{eqnarray}
\begin{split}
   \beta_{\mathrm{det}}(\bar{s}_{lm})=
   \int & f_{\mathrm{det}}(\Delta v',\theta,\phi,\vec{\gamma})\Pi(\Delta v',\theta,\phi|\bar{s}_{lm})\\
   & \times \Pi(\vec{\gamma})d\Delta v'd\theta d\phi d\vec{\gamma}.
   \end{split}
\end{eqnarray}
If we choose $\Pi_{\mathrm{sim}}(\Delta v',\theta,\phi,\vec{\gamma})=\pi(\Delta v')\pi(\theta)\pi(\phi)\Pi(\vec{\gamma})$, where $\pi(\Delta v'),\pi,(\theta),\pi(\phi)$ are the same as the ones defined in Eqs.\ \eqref{BH2}\ and \ \eqref{Bayes}, then priors in the denominator and numerator of the integrand in \eqref{BHpdet1} cancel out and we can define the marginalized fraction of detectable events (up to the factors that cancel out later):
\begin{eqnarray}
   f_{\mathrm{det}}^{\mathrm{marg}}(\Delta v',\theta,\phi)&\propto&\int p(\Delta v',\theta,\phi,\vec{\gamma}|det) d\vec{\gamma}.\label{BHpdet2}
\end{eqnarray}
As in the case of Eq.\ \eqref{BH3}, we have ignored a factor of $1/\pi(\Delta v')$ in Eq.\ \eqref{BHpdet2} for the same reason mentioned before. In terms of this marginalized fraction, $\beta_{\mathrm{det}}$ becomes
\begin{eqnarray}
    \begin{split}
   \beta_{\mathrm{det}}(\bar{s}_{lm})\propto\int& f_{\mathrm{det}}^{\mathrm{marg}}(\Delta v',\theta,\phi)\delta(\Delta v'-\Delta \nonumber v(\theta,\phi,\bar{s}_{lm}))\\
   & d\Delta v'd\theta d\phi.
   \label{BHsel2}\end{split}\\
\end{eqnarray}
To estimate $p(\Delta v',\theta,\phi,\vec{\gamma}|det)$ and hence $f^{\mathrm{marg}}_{\mathrm{det}}(\Delta v',\theta,\phi)$, we simulate a large number of events whose parameters are drawn from a broad distribution. We then inject the corresponding signals into detector noise realizations and record their SNRs and arrival times. We then apply our selection criteria to find which of these simulated events are detectable given our criteria and estimate $p(\Delta v',\theta,\phi,\vec{\gamma}|det)$. The estimation schemes will depend on which of the two approximations referred to before are used to smooth out the delta function integral and are hence described in more detail in the corresponding subsections below.

The priors we use to draw the simulated events are truncated power law in the primary mass and mass ratio, uniform in spin, sky position, orientation, comoving volume, geocentric time, and speed of GWs. Particularly, for each observing run, the mass distributions are chosen to be consistent with corresponding population analyses performed by the LVC such that the distributions used have support in regions of the mass space where the events being analyzed are found. For O2, we choose $p(m_1)\propto m_1^{-1.6}$, $ m_1\in(7.9M_{\odot},42M_{\odot})$, and $p(q)\propto q^{6.7}$, where $q=\frac{m_2}{m_1}$, which is consistent with Ref.\ \cite{Abbott2019-pop} and is identical to the mass distributions used for similar selection function computations \cite{Payne2020}. For O3, we choose $p(m_1)\propto m_1^{-1.6}$, $ m_1\in (7M_{\odot},80M_{\odot})$ and $p(q)\propto q^{6.7}$ , which is broad enough for the O3 events as evident from Ref.\ \cite{GWTC-3pop}. In the next two subsections, we describe the details of our smoothing approximations and the computation $\beta_{\mathrm{det}}$ in each approximation scheme.

\subsubsection{Narrow Gaussian method} 
The approach introduced here involves estimating the delta function in Eq.\ \rf{BH3} as a narrow Gaussian distribution. For each sample with measured speed difference $\Delta v'$ and sky location $\theta$ and $\phi$, we construct a Gaussian distribution for the random variable $\Delta v'-\Delta v(\bar{s}_{lm},\theta,\phi)$ with mean zero and standard deviation $\sigma$. Thus, Eq.\ \rf{BHsel1} becomes
\begin{equation}
    \begin{split}
        L(\bar{s}_{lm})=  \fr{1}{\beta_{\rm det}^{N}(\bar{s}_{lm})} &\prod_{i\in \{\rm events\}}^{N} \int  p(\Delta v',\theta,\phi|d_i)\\
        &\times\mathcal{N}(\Delta v'-\Delta v(\bar{s}_{lm},\theta,\phi))  d\Delta v' d\theta d\phi, \label{LikeN}
    \end{split}
\end{equation}
where $\mathcal{N}$ represents Gaussian distributions.  Similarly, we can also apply the narrow Gaussian approximation to the computation of $\beta_{\rm det}$ in Eq.\ \eqref{BHsel2}:
\begin{eqnarray}
\nonumber
     \beta_{\rm det}(\bar{s}_{lm}) &=& \int f_{\rm det}^{\rm marg}(\Delta v', \theta,  \phi) \mathcal{N}(\Delta v'-\Delta v(\bar{s}_{lm},\theta,\phi))  \\ && \times d\Delta v' d\theta d\phi. \label{LikeNdet}
\end{eqnarray}
Since $\mathcal{N}$ is a smooth function of its arguments we can evaluate the two integrals in Eqs.~\eqref{LikeN} and \eqref{LikeNdet} as a Monte Carlo sum over samples drawn from $p(\Delta v',\theta,\phi|d_i)$ and $f_{\rm det}^{\rm marg}(\Delta v', \theta,  \phi)$, respectively. Since we already have posterior samples drawn from $p(\Delta v',\theta,\phi|d_i)$ for each event during the $v_g$ inference described in Sec.\ \ref{sec:sog}, and since the samples drawn from $f_{\rm det}^{\rm marg}(\Delta v', \theta,  \phi)$ are the parameters of simulated events that survive our selection criteria,  we can compute the log-likelihood of $\bar{s}_{lm}$,
\begin{eqnarray}
         &&\ln L(\bar{s}_{lm})=
         \sum_{i\in\{\rm events\}}     \label{likeN2}\\ \nonumber && \phantom{ln}\ln{\fr{\sum_{\{j \}}\mathcal{N}(\Delta v'_j-\Delta v(\bar{s}_{lm},\theta_j,\phi_j))}{\sum_{k}\mathcal{N}(\Delta v'_k-\Delta v(\bar{s}_{lm},\theta_k,\phi_k)}},
\end{eqnarray}
where the sum in the numerator is over posterior samples corresponding to the $i$th event while the one in the denominator is over detectable samples. After choosing a width $\sigma$ for our Gaussian $\mathcal{N}$ appropriately, we can thus use Eq.~\eqref{likeN2} for fast evaluation of the log-likelihood $\ln{L(\bar{s}_{lm})}$ as a numerical function of the SME coefficients. Hence, we can use MCMC algorithms to draw samples from $\ln{L(\bar{s}_{lm})}$ and interpret the quantile ranges of said samples as Bayesian credible intervals of the SME coefficients given GW data.

To determine the appropriate width of our Gaussian distribution $\sigma$, we consider the effect of varying its size. Because the Gaussian distribution is an estimation of the delta distribution, theoretically, as the size of $\sigma$ decreases, the approximation should be more accurate. However, because we sample the log-likelihood with a MCMC algorithm, we encounter numerical difficulties when the $\sigma$ is too small. Thus our choice of $\sigma$ has to be tuned in accordance with how the MCMC is implemented numerically.

In the MCMC process, the walkers only make use of local information at each step. Thus, it is possible for walkers to be trapped inside islands of high likelihood. This is what happens when $\sigma$ is set too small. Since most samples have a high likelihood around zero, walkers can explore freely the region near zero. However, at more peripheral locations in the parameter space, the peaks are usually scattered. Thus, when $\sigma$ is too small, these peripheral samples form isolated islands of high likelihood. In this case, the walkers will not be able to explore these isolated islands, resulting in false small constraints. 
On the other hand, as $\sigma$ gets larger, our approximation becomes less accurate and distributions are artificially broadened.  Therefore, we aim to find a $\sigma$ such that it is big enough for the walkers to explore the sample space fully and small enough such that it gives us useful results. 

\begin{figure*}[htp]
    \centering
    \includegraphics[width=1\textwidth]{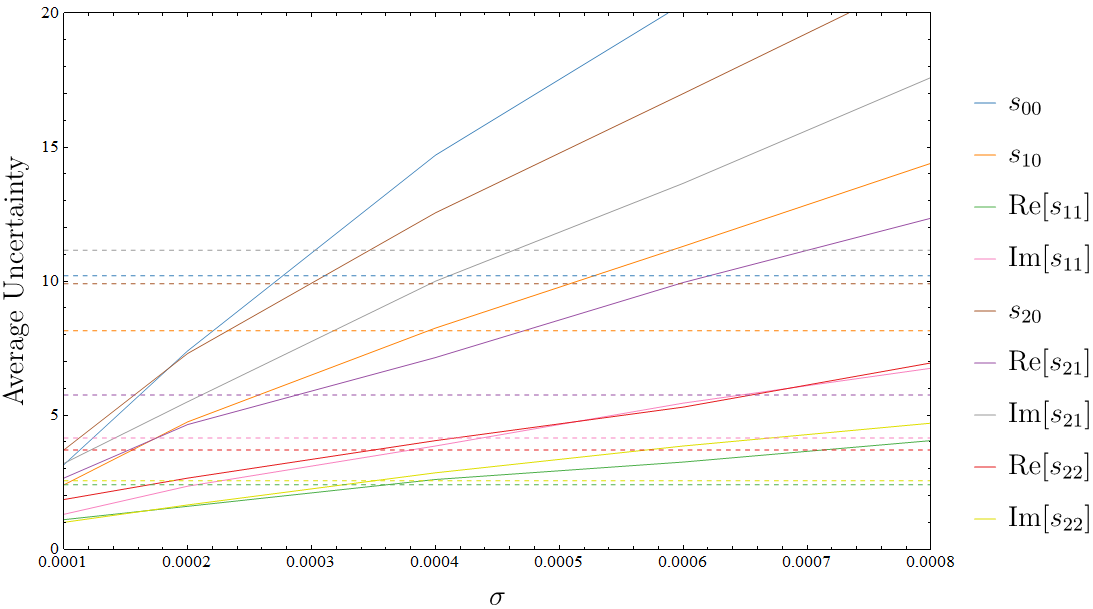}
    \caption{A portion of the superimposed plot of average uncertainties produced by the random draw method (dashed lines) and the Bayesian method (solid lines). From this plot, we can see $\sigma=0.0005$ is a potential choice for $\sigma$ because every solid line is marginally above the corresponding dashed line of the same color.}
    \label{fig:SvsSigma_combined}
\end{figure*}

One way to determine the appropriate value of $\sigma$ is by applying both the random draw method and the narrow Gaussian method on the same set of data and comparing the results.  We divide our list of events into subsets of nine events and apply both methods to each subset using various sizes of $\sigma$. The solid lines in Fig.\ \ref{fig:SvsSigma_combined} represent the average uncertainty of the resultant $\bar{s}_{lm}$'s against $\sigma$ for the nine O3 events from this paper with the smallest sky areas. The average uncertainty is calculated by taking the average of the absolute value of the upper and lower one-standard-deviation value for each $\bar{s}_{lm}$.  For the same sets of events, the random draw method produces uncertainties on the order of $10^0$--$10^1$, which corresponds to dashed lines in Fig.\ \ref{fig:SvsSigma_combined}. To select a suitable $\sigma$, we use superimposed plots such as Fig.\ \ref{fig:SvsSigma_combined} to select a $\sigma$  such that each uncertainty produced by the narrow Gaussian method is marginally larger than its counterpart from the random draw method.  In the example shown, $\sigma=0.0005$ is a good choice because every solid line lies marginally above the corresponding dashed line in the same color, which means that the choice $\sigma$ does not artificially tighten the constraints. We perform such analysis for every subset of events and produce a ``good $\sigma$." We pick the largest of such ``good $\sigma$'s" as the final choice. Even though this final $\sigma$ produces wider constraints for each subset of events as compared to the random draw method, because the narrow Gaussian method incorporates information from more than nine events, ideally we could still produce tighter constraints than the random draw method.

\begin{figure*}[!htp]
    \centering
    \includegraphics[width=1\textwidth]{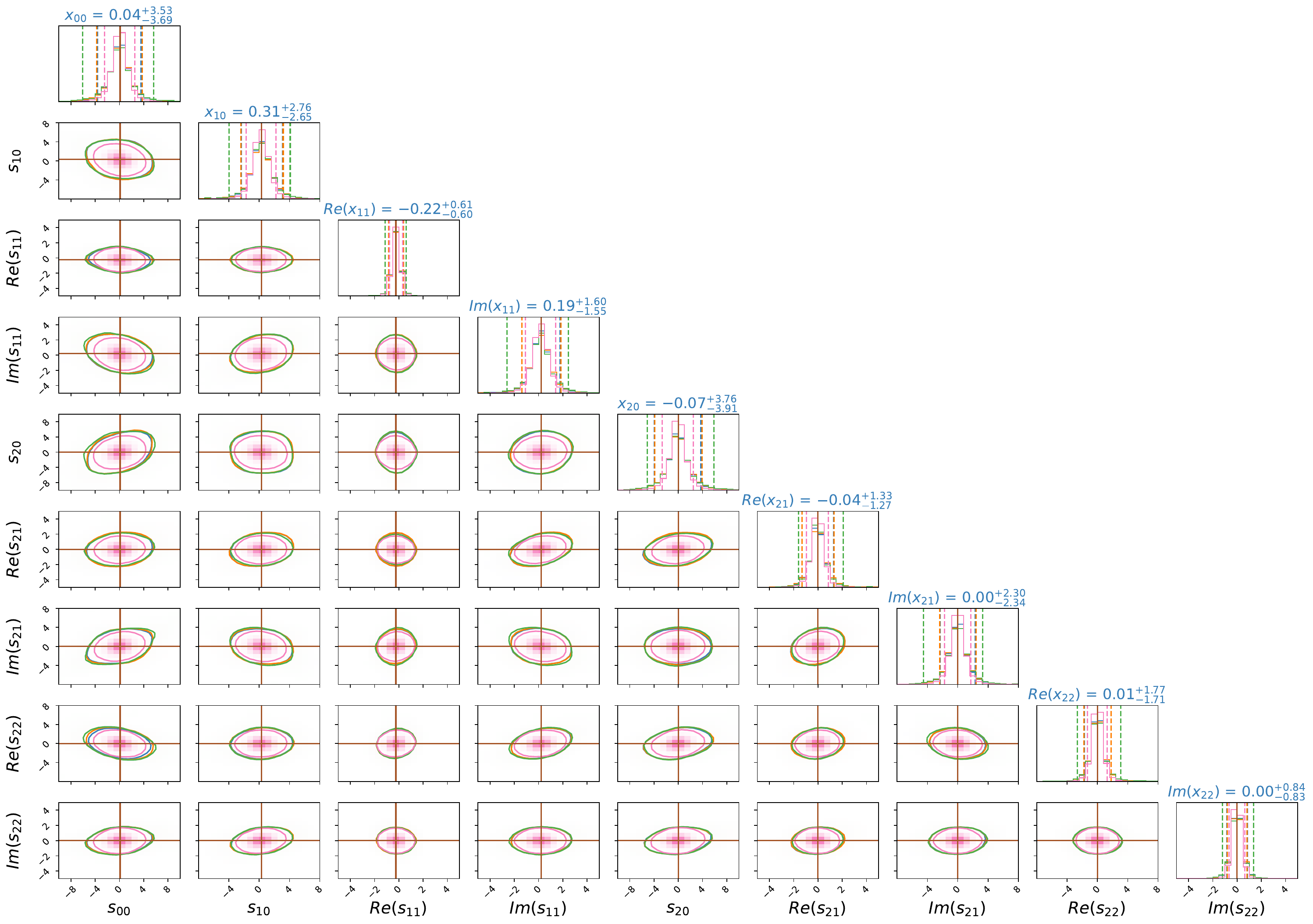}
    \caption{Distribution of all nine $\bar{s}_{lm}$ for the mock data. The exact Bayesian, KDE, narrow Gaussian, and SVD chi-square methods are color coded blue, orange, green, and magenta respectively. The mock true values of $\bar{s}_{lm}$, which were chosen to be either equal to or very close to zero, are marked by orange lines. Numbers above the plots show median values with 90\% equal tail credible intervals. The displayed constraints were obtained from the inferred $\bar{s}_{lm}$ samples using the \textsc{Python} package \textsc{corner}~\cite{corner}. Note that these mock events are ``zero noise" in the sense that the mean values of the Gaussian in Eq.\ \eqref{ExactBayesian1} are chosen to be equal to the true values and not perturbed by another random draw.}
    \label{fig:corner-mdc}
\end{figure*}

Another procedure for determining $\sigma$ is to use information directly from the distribution of $\Delta v'$ samples for each event. One specific procedure is to sort the list of $\Delta v'$ samples and compute the average difference between adjacent values. Choosing $\sigma$ as half this value produces results that align well with the prior method for the specific distributions tested. That is, the value of $\sigma$ for a given event is half of the average of the differences between adjacent $\Delta v'$ samples for that event. This procedure is used in the MDC shown in Fig. \ref{fig:corner-mdc}, and has the advantage of not needing to construct plots to determine $\sigma$. However, there still is subjectivity in choosing what fraction of the average to use. Moreover, the distribution and quantity of samples will impact the value of $\sigma$. Both considerations will affect the final credible intervals for $\bar{s}_{lm}$.

This method performs much better than the SVD-assisted random draw in the MDC performed in Sec.\ \ref{sec:mdc}. However, we note that both processes for choosing $\sigma$ involve a significant amount of user-controlled fine-tuning and can potentially lead to an over-/underestimate of measurement uncertainties of the SME coefficients. For these reasons, we do not choose this method for our final results. We instead choose a different smoothing approximation to the Bayesian method by means of KDE, which can be shown to produce either equally or more accurate results, while requiring almost no user-controlled fine-tuning.

\subsubsection{KDE methods}

In this section we outline a different approach from the one in the previous section, to perform Bayesian hierarchical inference of the SME coefficients from GW data. In this method, instead of smearing out the delta function in $\Pi(\Delta v',\theta, \phi|\bar{s}_{lm})$ with a Gaussian, we approximate the marginalized posterior of $\Delta v, \theta, \phi$ given GW data, for each event, as a fast evaluating function of these quantities, from their single event parameter estimation samples via Gaussian KDE. 

The KDE approximation of the posterior is constructed by fitting a multivariate Gaussian around each posterior sample and then writing the estimate of the posterior as a sum of these individual Gaussians. The covariance matrix of each of the Gaussians is approximated from the sample covariance matrix of the posterior samples themselves up to a constant of proportionality. The constant of proportionality is known as the bandwidth of the estimator and is computed, under reasonable assumptions regarding the true distribution being estimated(see Ref.~\cite{Scott1992}). We use SciPy's Gaussian KDE algorithm to obtain our estimate of the marginalized posterior as a fast evaluating function $p_{{\rm KDE},\;i}(\Delta v'=\Delta v(\theta,\phi,\bar{s}_{lm}),\theta,\phi)$ of the relevant parameters \cite{2020SciPy-NMeth}. We then perform the $\Delta v'$ integral of Eq.\ \eqref{BH3} analytically using the delta function and compute the remaining two integrals (over $\theta,\phi$) numerically using the trapezoidal rule. We loop over multiple events by multiplying the value of the integral obtained using the KDE corresponding to each event, to evaluate $L(\bar{s}_{lm})$ : 

\begin{equation}
    \begin{split}
    L(\bar{s}_{lm}) & \approx\\ & \prod_{i\epsilon\{events\}} \int p_{KDE,i}(\Delta v'=\Delta v(\theta,\phi,\bar{s}_{lm}), \theta, \phi) d\theta d\phi.
    \label{eq:KDE}
    \end{split}
\end{equation}

We then sample from it using the same MCMC method described in the previous section to constrain the SME coefficients. To incorporate selection effects in the KDE method, we estimate $f_{\rm det}^{\rm marg}(\Delta v',\theta, \phi|det)$ by performing a KDE on the samples of $(\Delta v',\theta, \phi)$ for which the simulated events are detectable given our detection criteria. By restricting our KDE to only these parameters and ignoring other parameters that characterize a simulated event, we effectively marginalize over those other parameters, thus implicitly performing the integral in Eq.\ \eqref{BHpdet2},
\begin{eqnarray}
    f_{\rm det}^{\rm marg}&\approx&  p_{\rm KDE,\; det}(\Delta v', \theta, \phi),
\end{eqnarray} 
where the subscript $det$ in $p_{KDE,det}$ represents the fact that this KDE was performed on only those samples for which the simulated events are detectable given our detection criteria. Substituting into Eq.\ \eqref{BHsel2}, we get
\begin{eqnarray}
    \beta_{\rm det}(\bar{s}_{lm}) &\approx & \int p_{\rm KDE,\;det}(\Delta v'=\Delta v(\theta,\phi,\bar{s}_lm), \theta, \phi)d\theta d\phi.\nonumber \\
    \label{selKDE}
\end{eqnarray}

We note that the KDE's bandwidth acts like a control parameter with potential room for user-controlled fine-tuning in the computation of its value, somewhat analogous to the $\sigma$ of the narrow Gaussian method. However, unlike the narrow Gaussian method where $\sigma$ can in principle be chosen to be anything, the bandwidth of the KDE is computed directly from the properties of the samples (such as the number of samples and dimensionality of the parameter space) under reasonable assumptions regarding the true density. Thus, the user's choice is restricted to a number of discrete such assumptions (for example, Scott's rule~\cite{Scott1992}, Silverman's rule~\cite{Silverman1986}, etc.). Furthermore, the effects of choosing a bandwidth on the estimated density (and hence the remainder of the inference) is limited in the sense that the covariance matrix of the Gaussians is determined from the samples themselves with the bandwidth only acting as a scaling parameter that is usually of order unity. This additionally restricts the effects of user-controlled fine-tuning on the inference as compared to the narrow Gaussian method wherein the width of the Gaussian that approximates the delta function is completely determined by the user's choice. A more detailed discussion of this comparison between the two methods in the context of the MDC can be found in Sec.~\ref{sec:mdc} . For our chosen bandwidth approximation scheme (Scott's rule)  the KDE method can be seen to perform extremely well in the MDC. For these reasons, we choose the KDE method for our final results on the SME constraints.

\begin{figure*}[htp]
    \centering
%    \twocolumn
    \includegraphics[width=1\textwidth]{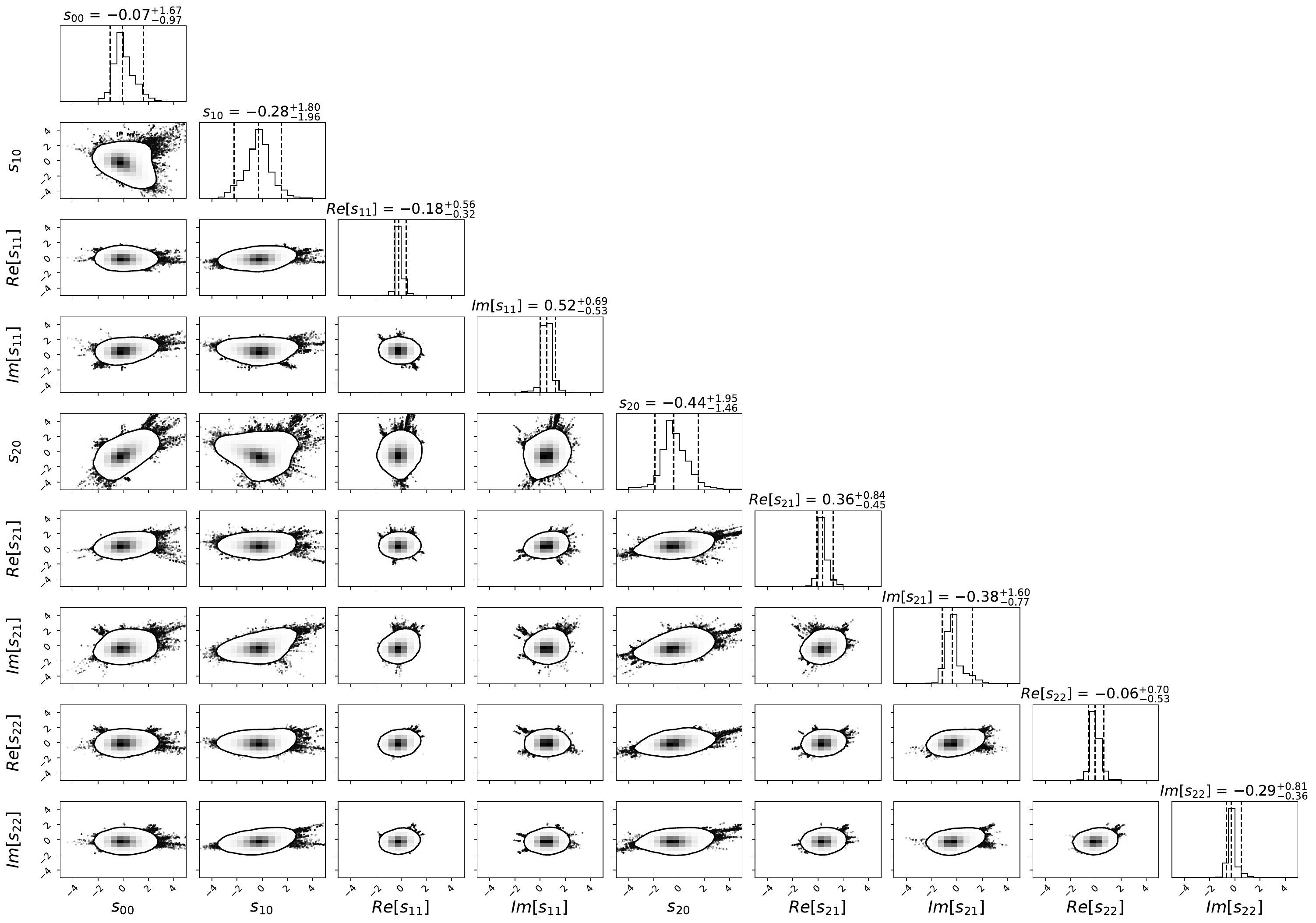}
    \caption{Distribution of all nine $\bar{s}_{lm}$ using the KDE method from the 24 chosen GW events in Tables \ref{table:O3a}, \ref{table:O3b}, and \ref{table:O12}. The numbers above the plots show median values with 90\% equal tail credible intervals. Each contour represents the two-dimensional 90\% credible interval marginalized over the other seven parameters while the discrete points represent posterior samples that lie outside of the 90\% contour. The grayscale inside the contour represents the two-dimensional marginalized posterior density in the form of a histogram. The displayed constraints were obtained from the inferred $\bar{s}_{lm}$ samples 
 using the \textsc{Python} package \textsc{corner}~\cite{corner}. }
    \label{fig:corner}
\end{figure*}

 We present the results of the KDE method upon its use in the analysis of the events marked * listed in tables I through III (except for GW170817 for which we do not use the fixed posteriors due to the inability of the KDE to estimate very narrow densities) in Fig.\ \ref{fig:corner} as our final result.

\subsection{Discussion}
The random draw method, which was originally presented in Ref.~\cite{O12sog}, is very time efficient. However, the number of events that can be used in the analysis is limited by the number of $\bar{s}_{lm}$ to be estimated. Hence, it is only useful in the scenario wherein we have exactly the same number of events available to be used in the analysis as the number of $\bar{s}_{lm}$ coefficients being simultaneously measured. On the other hand, unlike any of the following methods, this method does not involve any approximation of the delta distribution. Thus, this method can be used as a quick feasibility test. For example, we used this method to estimate the size of $\sigma$ for the narrow Gaussian approach. 

With the use of SVD, we were able to take data from a larger set of events. However, the presence of one ``bad" event, i.e., a low-significance event with biased posteriors, can disturb the entire analysis since all events are given equal weights. With the SVD chi-square method, this problem is solved by weighting each event with its uncertainty in the $v_g$ measurement. However, this leads to artificially narrower bounds with ambiguity in the meaning of those bounds.

The Bayesian methods are free of all aforementioned pathologies that plague the other methods. They can efficiently handle a large number of events and is unaffected by a small number of bad events if any. Furthermore, the Bayesian credible intervals have the clear and unambiguous meaning of being regions of the parameter space that contain the true value of said parameters with a certain posterior probability given data.

Among the two approximate Bayesian techniques described in this paper, the narrow Gaussian method has the following issue. The process to find $\sigma$ is cumbersome and somewhat subjective as it involves partitioning the set of events into subsets and estimating $\sigma$ from plots, or developing an algorithm that needs to be compared to an independent method. On the other hand, the KDE method has less user-controlled fine-tuning than the narrow Gaussian method as it estimates its control parameter, i.e. the bandwidth, quantitatively from properties of the posterior samples themselves, with its variation having a much more restrictive effect on the estimated density. Hence, we claim that the Bayesian analysis implemented by the KDE method produces the most trustworthy measurements of the SME coefficients.

\subsection{Mock data challenge and comparison of methods}
\label{sec:mdc}
In this section, we describe the MDC that was set up to compare the different methods of SME measurements from GW data in order to verify our claims regarding them that were made in the previous section. To construct the MDC, we choose a fiducial value of the SME coefficients as their true values, say $\bar{s}_{lm,tr}$ and generate data for 15 mock events. The true sky positions of the mock events are chosen to be the mean values of the $(\theta,\phi)$ samples of the real events. 

We choose 15 of the ``best" real events, i.e.\ the ones with the most precise sky localizations and $v_g$ measurements, to be represented by our mock events in the MDC. We then calculate the true value of $\Delta v$ for each mock event from the true values of their sky positions and those of the SME coefficients. We then generate mock posterior samples of $(\theta,\phi,\Delta v)$ by adding uncorrelated Gaussian fluctuations to the true sky positions and true $\Delta v$ values. The width of the fluctuations for each mock event is chosen to be the standard deviations of the posterior samples of the corresponding real event. 

This allows us to create a controlled numerical experiment wherein we know the true answer. For Gaussian distributions of $\theta,\phi,\Delta v$ about known true values, one can write down the exact functional form of the likelihood of these parameters given mock data,
\begin{eqnarray}
     L_{\rm mdc}(d_i|\Delta v', \theta, \phi)&=&\frac{1}{(2\pi)^{3/2}\sigma_{\theta,i}\sigma_{\phi,i}\sigma_{\Delta v,i}}\exp -\frac{1}{2}\{\frac{(\theta-\theta_{tr,i})^2}{\sigma_{\theta,i}^2}\nonumber\\
     &~&+\frac{(\phi-\phi_{tr,i})^2}{\sigma_{\phi,i}^2}+\frac{(\Delta v'-\Delta v_{tr,i})^2}{\sigma_{\Delta v,i}^2}\},\label{ExactBayesian1}
\end{eqnarray}
where $(\theta_{tr,i},\phi_{tr,i})$ are the true values of the sky positions of the $i$th mock event; $\sigma_{\theta,i},\sigma_{\phi,i},\sigma_{\Delta v,i}$ are the widths of the Gaussian fluctuations used to generate the mock posterior samples of the $i$th mock event; and $\Delta v_{tr,i}=\Delta v(\bar{s}_{lm,tr},\theta_{tr,i},\phi_{tr,i})$. Knowledge of these quantities allows us to exactly write down and evaluate Eq.\ \eqref{ExactBayesian1} as a function of $\theta,\phi$ without any smoothing approximations. 

We can then substitute $L_{\rm mdc}(d_i|\Delta v',\theta,\phi)$ in place of $L(d_i|\Delta v',\theta,\phi)$ in Eq.\ \eqref{BH1} and carry out the integral numerically to obtain the ``exact Bayesian" likelihood of our mock data given the SME coefficients:
\begin{eqnarray}
     L_{\rm mdc}(\bar{s}_{lm})&=&\prod_i \frac{1}{(2\pi)^{3/2}\sigma_{\theta,i}\sigma_{\phi,i}\sigma_{\Delta v,i}}\int\exp -\frac{1}{2}\{\frac{(\theta-\theta_{tr,i})^2}{\sigma_{\theta,i}^2}\nonumber\\
     &~&+\frac{(\phi-\phi_{tr,i})^2}{\sigma_{\phi,i}^2} \label{ExactBayesian2} \\
     &~&+\frac{(\Delta v(\bar{s}_{lm},\theta,\phi)-\Delta v(\bar{s}_{lm,tr},\theta_{tr,i},\phi_{tr,i}))^2}{\sigma_{\Delta v,i}^2}\}d\theta d\phi. 
     \nonumber
\end{eqnarray}
We can then sample the likelihood in Eq.\ \eqref{ExactBayesian2}, after applying suitable priors on $\bar{s}_{lm}$, using the MCMC techniques described above and obtain what can be thought of as the true posterior distribution of the SME coefficients given the mock data. We can then compute the constraints obtained from the approximate methods being applied to the mock posterior samples and compare those results with the true posterior.

The results of this comparison are displayed in Fig.~\ref{fig:corner-mdc}. We can see that the KDE method agrees remarkably well with the exact Bayesian method, while the narrow Gaussian method deviates from it slightly for some coefficients. We note that a different choice of $\sigma$ for the narrow Gaussian method leading to better agreement with the exact Bayesian result is possible. However, we conclude that the KDE method's agreement with the exact Bayesian method, independent of any external fine-tuning, justifies its use on the real data for producing our final result. We also note that our claim regarding the SVD chi-square method producing artificially narrower bounds is also verified by this comparison.

\subsection{Final results}

With the 24 chosen GW events in Tables \ref{table:O3a}, \ref{table:O3b}, and \ref{table:O12}, we are able to constrain all nine $\bar{s}_{lm}$ coefficients. We obtain the results shown in Fig. \ref{fig:corner}.

Note that the measurements of $\bar{s}_{lm}$ shown in Fig.\ref{fig:corner} are consistent with zero. Given that zero lies within the 90\% credible intervals for all coefficients, we consider these results to be consistent with existing constraints on $\bar{s}_{lm}$ \cite{data}. 

These limits are considerably weaker than some found in the literature. However, they are valuable as independent tests. Moreover, this is also the first attempt to simultaneously constrain all $\sb_{lm}$ using GW measurements, thus putting direct limits on the full potential anisotropy of the speed of GWs. Additionally, this method can theoretically incorporate as many events as available and thus improve in precision as additional high-quality events become available.

\section{Conclusion}
\label{sec:conclusion}

In our study, we selected 52 high-SNR gravitational-wave events that were detected by at least two detectors from the first three observing runs of Advanced LIGO and Advanced Virgo. We used \textsc{\texttt{lalinference\_nest}} and \textsc{\texttt{lalinference\_mcmc}} to construct posterior distributions of the speed of GWs for each event. We found the 90\% credible interval of the combined $v_g$ posterior distribution to be $0.99^{+0.01}_{-0.02}c$. This interval is narrower than the similar one constructed with O1 and O2 events in previous studies, suggesting a more precise measurement of $v_g$ \cite{O12sog}. However, even with the inclusion of a high-SNR BNS event GW170817 with its pinpoint sky localization, we were only able to narrow the 90\% credible interval to $0.99^{+0.01}_{-0.02}c$. We then explored multiple methods of extracting SME constraints from $v_g$-like data.  Based on the conclusions of that investigation,
we used hierarchical Bayesian inference implemented with KDE methods to simultaneously constrain all nine coefficients for Lorentz violation in the SME framework. The resultant constraints did not exhibit evidence for Lorentz violation. 
We are optimistic about the possibility of further improvements in speed of gravity and associated Lorentz violation measurements in the future.  The search is likely to be aided by the combination of additional detectors at additional locations around the Earth and by the possibility of combining results achieved by the methods presented here with those from additional multimessenger events.

\appendix
\section{Robustness against choice of waveform approximants}

\begin{figure*}[htp]
    \centering
%    \twocolumn
    \includegraphics[width=1\textwidth]{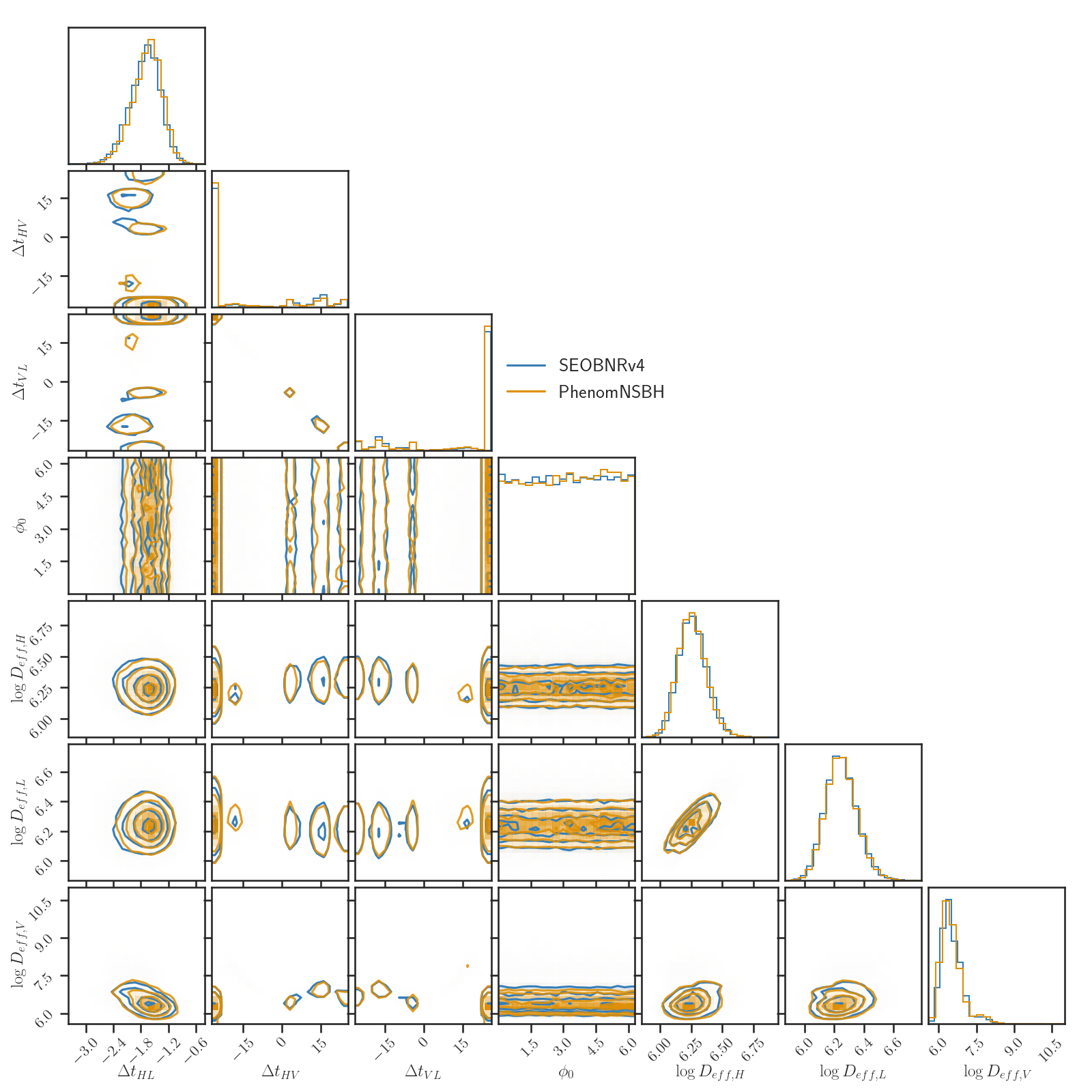}
    \caption{Insensitivity of observables to the choice of waveform approximants for a randomly selected O3b event~(GW200115).}
    \label{fig:S200115j}
\end{figure*}

In this section, we elaborate on the insensitivity of our results to the choice of waveform approximants. All measurements presented in this work rely on the posterior samples of $v_g,\theta,\phi$ that are obtained from single-event parameter estimation runs. Samples of these parameters are implicitly computed from those of direct observables such as the time delay between the arrival of signals at different detectors~($\Delta t_{HL},\Delta t_{HV}, \Delta t_{LV}$), the effective distance to the source as measured by each detector~($D_{eff,H},D_{eff,L},D_{eff,V}$), and the coalescence phase~($\phi_0$). In Fig.~\ref{fig:S200115j}, we show that parameter estimation results for such observables are fully consistent between different choices of the waveform approximant for a randomly chosen event in O3 (GW200115). The samples are obtained from LVK's public data release. We note that fixing $v_g=c$ (as was done for the publicly released samples) or allowing it to vary does not affect these observables since $v_g$ is a derived quantity while the observables are directly measurable.

\section*{ACKNOWLEDGMENTS}
\label{sec:ack}

This work was supported by NSF Grants No.~PHY-1806990, No.~PHY-1912649, and No.~PHY-2207728. This material is based upon work supported by NSF's LIGO Laboratory which is a major facility fully funded by the National Science Foundation. The analysis in this work used the LALSuite software library \cite{lalsuite}. We thank LIGO and Virgo Collaboration for providing the data from the first, second, and third observing runs. The authors are grateful for computational resources provided by the LIGO Laboratory and supported by National Science Foundation Grants No.~PHY-0757058 and No.~PHY-0823459.
We thank Nathan Johnson-McDaniel for helpful comments during a careful review of the manuscript.

This research has made use of data or software obtained from the Gravitational Wave Open Science Center (gwosc.org), a service of LIGO Laboratory, the LIGO Scientific Collaboration, the Virgo Collaboration, and KAGRA. LIGO Laboratory and Advanced LIGO are funded by the United States National Science Foundation (NSF) as well as the Science and Technology Facilities Council (STFC) of the United Kingdom, the Max-Planck-Society (MPS), and the State of Niedersachsen/Germany for support of the construction of Advanced LIGO and construction and operation of the GEO600 detector. Additional support for Advanced LIGO was provided by the Australian Research Council. Virgo is funded, through the European Gravitational Observatory (EGO), by the French Centre National de Recherche Scientifique (CNRS), the Italian Istituto Nazionale di Fisica Nucleare (INFN) and the Dutch Nikhef, with contributions by institutions from Belgium, Germany, Greece, Hungary, Ireland, Japan, Monaco, Poland, Portugal, Spain. KAGRA is supported by Ministry of Education, Culture, Sports, Science and Technology (MEXT), Japan Society for the Promotion of Science (JSPS) in Japan; National Research Foundation (NRF) and Ministry of Science and ICT (MSIT) in Korea; Academia Sinica (AS) and National Science and Technology Council (NSTC) in Taiwan.

\bibliography{main}
\end{document}